\documentclass[twocolumn,prb,aps,showpacs]{revtex4}
\pdfoutput=1
\usepackage[latin1]{inputenc}
\usepackage{amsmath,amsbsy}
\usepackage{amssymb}
\usepackage[hidelinks]{hyperref}
\usepackage{tikz}
\usepackage{subfigure}
\usetikzlibrary{arrows, decorations.markings}
\usepackage{rotating}
\usepackage{color}
\usepackage{graphicx}

\newcommand \cee {\ensuremath{\hat c^{\vphantom\dagger}}}
\newcommand \cdag {\ensuremath{\hat c^\dagger}}
\newcommand \eff {\ensuremath{\hat f^{\vphantom\dagger}}}
\newcommand \fdag {\ensuremath{\hat f^\dagger}}
\newcommand \iw {\ensuremath{\mathrm i\omega}}

\newcommand \Eq [1] {Eq.~\eqref{eq:#1}}

\begin{document}
\title{Mott-Hubbard transition in the mass-imbalanced Hubbard model}
\author{Marie-Therese Philipp, Markus Wallerberger, Patrik Gunacker, and  Karsten Held}
\affiliation{Institute of Solid State Physics,  TU Wien, A-1040 Vienna, Austria}

\begin{abstract}
The  mass-imbalanced Hubbard model represents a continuous evolution from  the Hubbard to the Falicov-Kimball model. We employ  dynamical mean field theory and study  the paramagnetic
 metal-insulator transition, which has a very different nature for the  two limiting models. Our results indicate that the
 metal-insulator transition rather resembles that of the Hubbard model  as soon as a tiny hopping between the more localized fermions is switched on. 
At low temperatures we observe  a first-order metal-insulator transition and
a three peak structure. The width of the central peak is the same for the
more and less mobile fermions when approaching the phase transition, which agrees with our expectation of a common Kondo temperature and phase transition for the two species.

\end{abstract}
\maketitle
\section{Introduction} \label{sec:Introduction}
\begin{figure}[t]
  \includegraphics[width=\columnwidth]{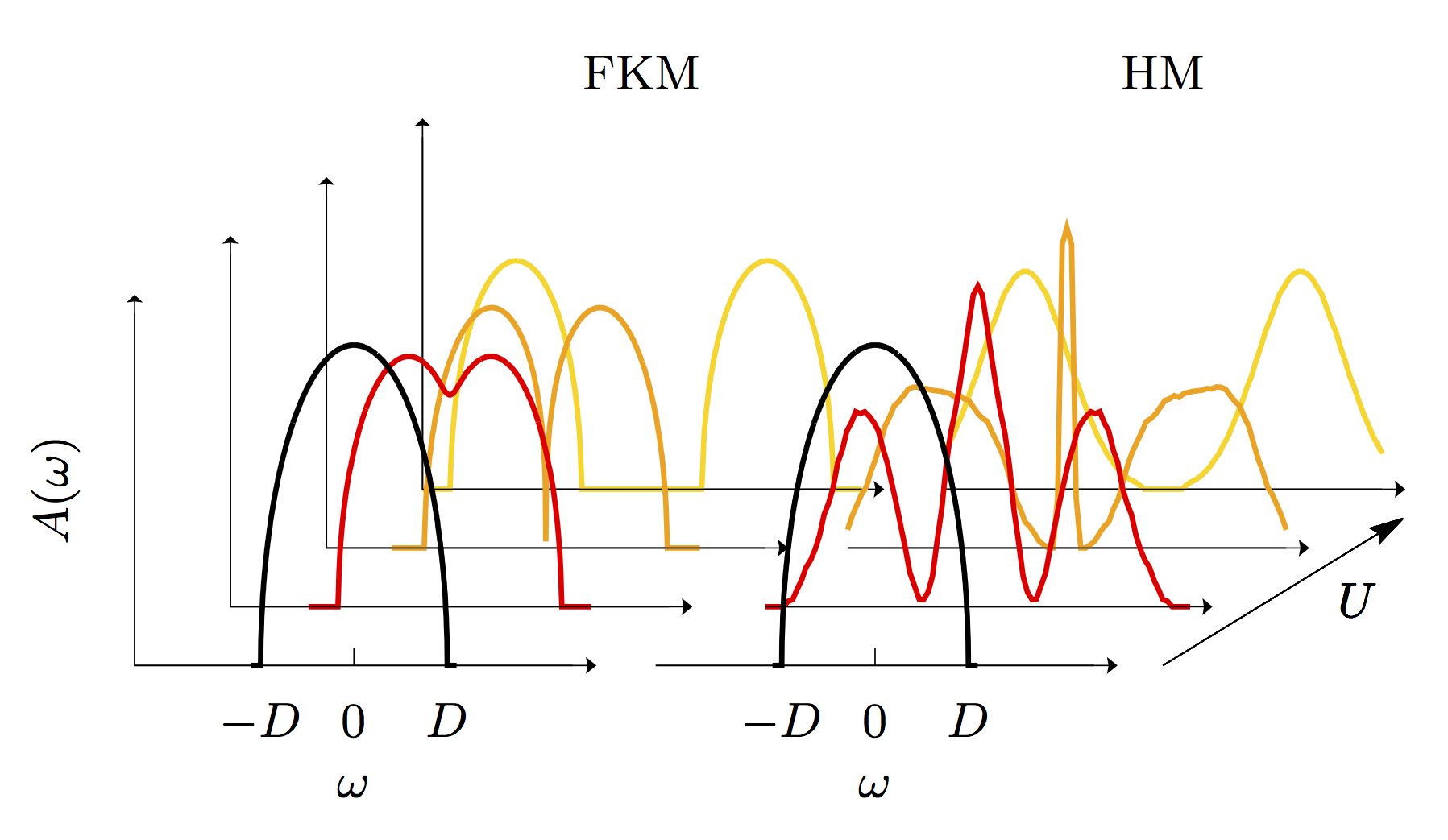}
  \caption{\label{fig0:scheme} Schematic view of the Mott-Hubbard metal-insulator transition in the paramagnetic phase of the Falicov-Kimball model (left) and Hubbard model (right). The spectral weight $A(\omega)$ at energy $\omega$ is shown along a qualitative interaction-axis $U$ taking into account the shift of the metal-insulator transition.\cite{footnoteschematicMIT} 
	      }
\end{figure}

Strongly correlated electron systems give raise to a plethora of fascinating physics,
and the  Mott-Hubbard metal-insulator transition\cite{Gebhard97a,Imada1998} is one of the 
 prime examples. Here, electronic correlations split the non-interacting band so that an insulating state develops,
  even for an odd number of electrons per site and even without symmetry breaking in the paramagnetic phase.
The arguably first satisfactory modeling of this transition was given by Hubbard in the so-called Hubbard-III 
approximation.\cite{Hubbard63} This is a decoupling of the equation of motion for the Green function
 based on an alloy analogy, which  assumes mobile electrons moving in a lattice of immobile electrons. 
Physically this solution yields separated Hubbard bands at $\pm U/2$
for large interactions $U$, and a continuous metal-insulator transition  at a critical interaction strength $U_c$, where both 
Hubbard bands touch each other, see Fig.\ \ref{fig0:scheme} (left). Upon further reducing $U$, the pseudogap gets filled 
and eventually the non-interacting density of states (DOS) is recovered.\cite{vanDongen97}
While the Hubbard-III approximation was designed as an approximation for solving the Hubbard model (HM),
it already bares the main ingredients of the Falicov-Kimball model (FKM):\cite{Falicov69} mobile and immobile electrons that are coupled through an interaction $U$. Because of this resemblance,  Hubbard\cite{Hubbard63} is sometimes given credit for
inventing the FKM. 

In particular thanks to insights from dynamical mean field theory (DMFT), \cite{Georges92a,Georges96a} we nowadays know that the
Mott-Hubbard transition in the Hubbard model is more intricate. First of all, it is first order for temperatures  below a critical point, as in the van der Waals liquid-vapor transition. Second, when turning metallic, a quasiparticle resonance develops within the, still pronounced, gap. Consequently a three-peak structure emerges,  which only dissolves into a single peak at much smaller interaction strength, see Fig.\ \ref{fig0:scheme} (right). In contrast, the FKM shows the Hubbard-III type of metal-insulator transition  in DMFT \cite{vanDongen97,Freericks03}  as in Fig.\ \ref{fig0:scheme} (left).

Since these two metal-insulator transitions are genuinely different, we ask in this paper: What is the nature of the metal-insulator transition when continuously tuning our model from the FKM to the HM?

We address this question  by increasing the hopping amplitude of the less mobile fermions from zero (FKM) to the same amplitude as the mobile fermions (HM).  As one may construe the different hopping amplitude
as a mass difference, one can also perceive this intermediate model as a mass-imbalanced Hubbard model. While the paradigm system
for the Mott-Hubbard transition of the HM-type is V$_2$O$_3$,\cite{Hansmann2013} realizing the mass-imbalanced HM in the solid state is difficult. Usually, there is a Kramers spin-degeneracy and not a strong mass imbalance between the two spin species. However with the advent of cold atom systems, such mass imbalances in a two-component system can be realized readily,  either by having different atomic species\cite{Taglieber08} or by generating a  spin-dependent hopping through a magnetic field gradient.\cite{Jotzu15}

On the theoretical side, the antiferromagnetic order in the mass-imbalanced HM has been studied before by DMFT \cite{Sotnikov12} and 
quantum Monte Carlo simulation \cite{Liu15}, but to the best of  our knowledge not the paramagnetic Mott-Hubbard transition which is the topic of the present paper.
Let us mention however that the  paramagnetic metal-insulator transition has been addressed  quite intensively in the literature
for the multi-orbital Hubbard model which has  different bandwidths or masses for the two orbitals.\cite{Liebsch03b,Biermann05b,Koga04a,Arita05a,Knecht05a,Ferrero05a} Here an orbital-selective Mott transition was found, where first one band and subsequently the second band turns insulating with increasing $U$. An important and, as we will later see, crucial difference to our situation is however that  both orbitals consists of  two spin species each.

In the present paper we address the Mott-Hubbard metal-insulator transition in the mass-imbalanced Hubbard model, where both spin species have a different mass or hopping amplitude. This allows us to continuously tune the model from the FKM to the HM. Section \ref{Sec:model} introduces the model and the method employed: DMFT using continuous quantum Monte Carlo simulations as an impurity solver.
In Section \ref{Sec:results} we present the calculated phase diagram of the mass-imbalanced HM, the self-energy, spectral function, and double occupancy. In Section \ref{Sec:discussion}
we discuss the underlying physics of a common Kondo scale for the two fermionic species and support this by numerical data which
show that the width of the central quasiparticle peak converges towards the same value for the two different fermions when approaching the metal-insulator transition. Consequently, we observe a single metal-insulator transition and not a  spin-dependent one as might be expected from
the aforementioned studies on  multi-orbital Hubbard model. 
Finally,  Section \ref{Sec:conclusion} summarizes our results.

\section{Model and method}
\label{Sec:model} 

The Hamiltonian of the mass-imbalanced Hubbard model for one orbital can be written as:
\begin{equation}
  \begin{split}
   H &= -t_c \sum_{\langle ij\rangle} ( \cdag_i\cee_j + \cdag_j\cee_i)
            - t_f \sum_{\langle ij\rangle} ( \fdag_i\eff_j + \fdag_j\eff_i) \\
            &\hphantom= + U \sum_i \hat n_{c,i} \hat n_{f,i},
  \end{split}
  \label{eq:ham}
\end{equation}
where $\cdag_i$ ($\fdag_i$) and $\cee_i$  ($\eff_i$) create and annihilate,
respectively, one of the two fermion species, while $\hat n_{c,i} = \cdag_i\cee_i$ and $\hat n_{f,i} = \fdag_i\eff_i$ are the corresponding occupancy operators at site $i$.
The hopping between next neighbors, 
$\langle ij\rangle$, is mediated by the corresponding hopping amplitudes, $t_c$ and $t_f$; and  it costs  the Hubbard interaction $U$  if there is  a $c$ and  $f$ fermion on the same site $i$.

In the following, we consider a Bethe lattice, i.e., a
semi-elliptic densities of states 
  $D_x(\omega) = \, \frac{2}{\pi D_x} 
                    \sqrt{ 1 - ( {\omega}/{D_x} )^2 }$
with half bandwidth $D_x\sim t_x$ for the two fermionic species $x=c,f$.
In the following, we set $D_c \equiv 2$ as our unit of energy and vary  the mass balance\cite{NoteMIFactor} $D_f/D_c = t_f/t_c$ between 0 and 1.

The two limits of mass imbalance are evident from \Eq{ham}: In the case $D_f/D_c = 0$, the 
$f$ fermions are truly frozen and we arrive at the FKM. On the other hand, 
if $D_f/D_c = 1$, we can identify $c$ and $f$ with spin-up and spin-down, respectively, and obtain the mass-balanced HM.

The mass-imbalanced HM (\ref{eq:ham}) can be  solved exactly using dynamical mean field theory (DMFT), which maps the lattice model
self-consistently on a single impurity Anderson model (SIAM).\cite{Georges96a}  We use the w2dynamics
package,\cite{Parragh12,Wallerberger16} which employs continuous time quantum Monte Carlo (CT-QMC) in the hybridization expansion\cite{Werner06,Gull11} to solve the auxiliary SIAM. In order to obtain spectral functions $A(\omega)$ in real frequency from the imaginary frequency DMFT Green's function $G(\iw_n)$, the maximum entropy method (MAXENT) is used.\cite{Jarrell96}

Both in the FKM and  HM, the metal-insulator transition is shadowed by an anti-ferromagnetic dome for dimension $d\ge 3$.  In order to be able to study the Mott transition, we hence enforce the paramagnetic solution and  half-filling of both, $c$ and $f$ fermions.  In the HM, these constraints can be enforced by explicitly symmetrizing over spins and setting the chemical potential to $\mu = U/2$, respectively. Away from this HM limit, we again enforce half-filling by fixing $\mu = U/2$ such that  the total number of fermions per site is $n=1$. However to ensure the paramagnetic solution at half-filling for both, $c$ and $f$, fermions individually, we symmetrize the hybridization functions in imaginary time: $\Delta(\tau)\overset{!}{=}\Delta(\beta-\tau)$.

\section{Results}
\label{Sec:results} 

As mentioned in Section \ref{sec:Introduction},
the physics of the Mott-Hubbard metal-insulator transition
is of first order in the HM ($D_f/D_c=1$), and hence accompanied 
by a hysteresis loop in the phase diagram. 
This hysteresis loop is obtained numerically as follows: 
Increasing $U$ by taking a DMFT solution of a smaller $U$ 
as a starting point for the next DMFT iteration, 
yields a metallic solution up to a
critical interaction $U_\text{c1}$. Above
$U_\text{c1}$, the metallic (M) 
solution is no longer stable and the system becomes 
Mott-insulating (I) instead.
Decreasing $U$ from here, yields an insulating solution
down to a second critical interaction $U_\text{c2}<U_\text{c1}$.
Hence, in the limit of the HM, where $c$ and
$f$ fermions are equally mobile, there is a coexistence 
region of the metallic
and the insulating phase for $U_\text{c2}<U<U_\text{c1}$.
The Mott-Hubbard metal-insulator transition in the HM has been investigated thoroughly in the 
literature, among others for the semi-elliptic Bethe DOS.\cite{Georges96a,Bluemer03}
In the FKM on the other hand no coexistence 
region is observed. The Mott-like
metal-insulator transition occurs 
temperature-independent exactly at $U_\text{c}=D_c$ in the FKM,
as is analytically known and well reviewed 
in Ref.\ \onlinecite{Freericks03},
also see Fig.\ \ref{fig0:scheme}.

\begin{figure}
  \includegraphics[width=\columnwidth]{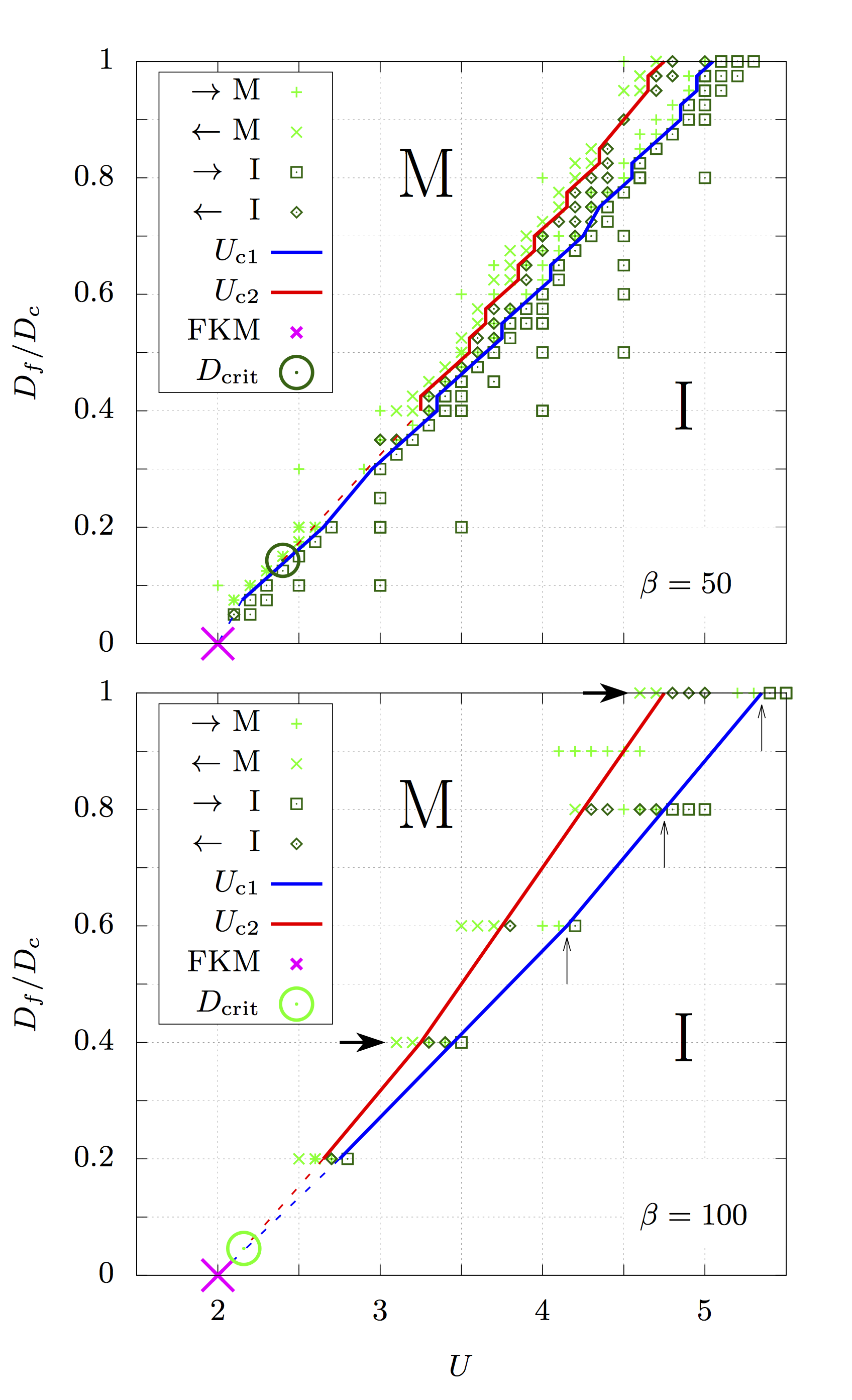}
          \caption{\label{fig1:PD}Phase diagram of the
          mass-imbalanced Hubbard model as a function of  
          interaction strength $U$ and mass imbalance $D_f/D_c$
          at $\beta=50$ (upper panel)
          and $\beta=100$ (lower panel); $D_c\equiv 2$ sets our unit of energy. 
          The critical
          interaction strengths $U_\text{c1}$ (blue line) has been obtained 
          by increasing $U$ ($\rightarrow$)
          and identifying up to which $U$ value the metallic solution
          ({M}, green plus) is still stable and from which $U$ value on 
          we get an insulating solution ({I}, green boxes).
          For decreasing $U$ ($\leftarrow$), $U_\text{c2}$ (red line) 
          marks the point where the insulating solution (green diamonds) 
          turns metallic (green crosses). The critical point where 
          the first order transition with coexistence region ends
          is extrapolated by hands of Fig.\ \ref{fig5:coex} and indicated 
          here by a green circle. The pink cross denotes the 
          analytical continuous phase transition for the FKM.
          }
\end{figure}

Fig.\ \ref{fig1:PD} shows the phase diagram  $D_f/D_c$ vs.\ $U$  of the
mass-imbalanced HM inbetween 
these two known limits at two inverse temperatures $\beta=50$ and $100$.  The first order coexistence region has been determined in the same way as described above for the HM: The four green symbols in Fig.\ \ref{fig1:PD} mark up to which point a metallic (M) or insulating (I) solution is found  upon increasing or decreasing $U$.
We observe a coexistence region and hence a first order transition in a wide range of mass imbalances. The coexistence 
region is increasing upon decreasing temperature to $\beta=100$, and the critical point where the first order transition ends (green circle, $D_\text{crit}$) is moving
towards the FKM limit $D_f/D_c=0$. To obtain the phase diagram, we  discriminate between metallic and insulting solution by means of the imaginary part of the
self-energy $\Sigma(\text{i}\omega_n)$. 

\begin{figure}
  \includegraphics[width=\columnwidth]{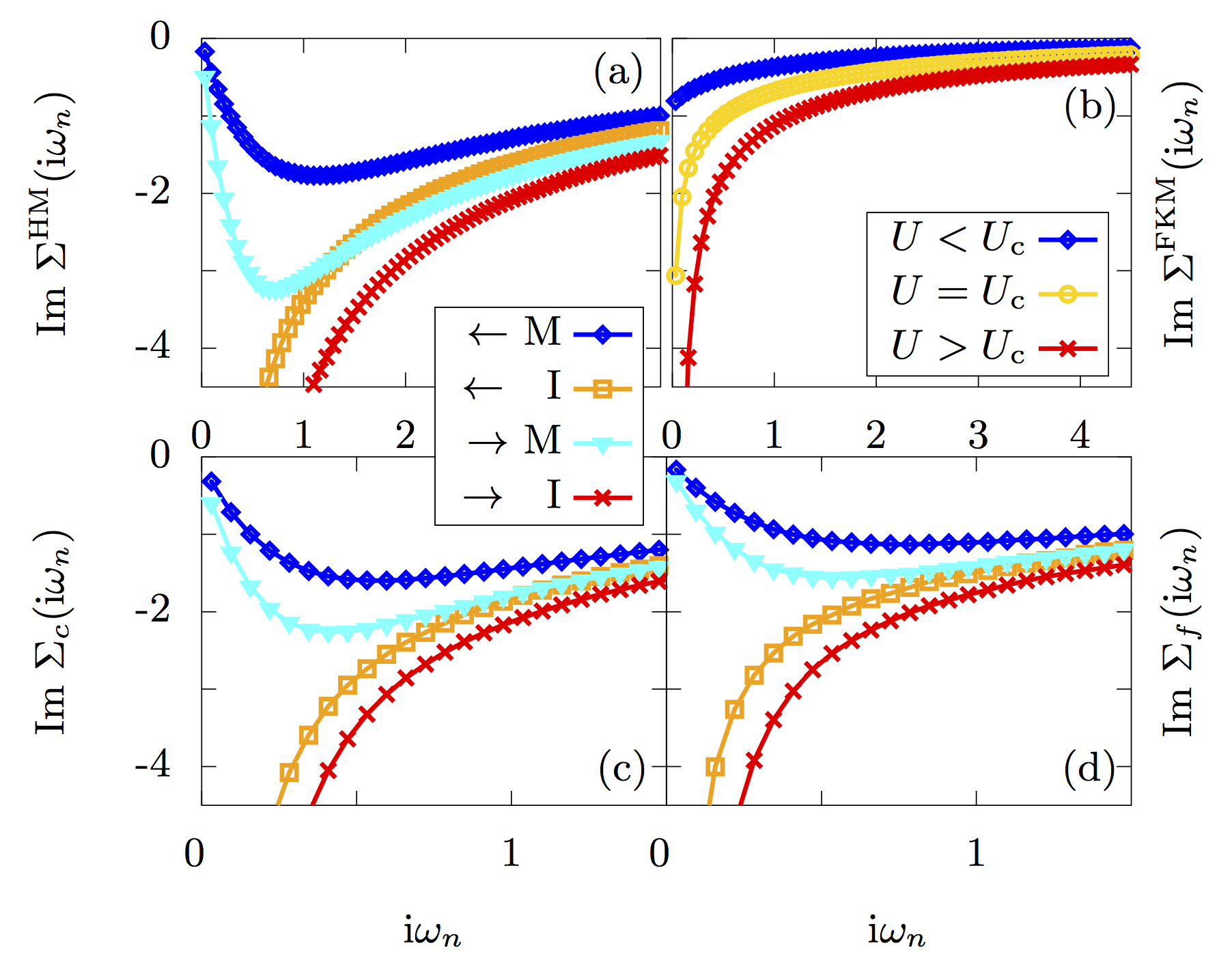}
	  \caption{\label{fig2:siw} Imaginary part of the 
      self-energy at $\beta=100$ as an indicator for a metallic and insulating solution.
      (a) HM with $D_f/D_c = 1$. 
      (b) FKM with $D_f/D_c = 0$.
      Lower panel:  Comparison of the $c$ fermions (c) and $f$ fermions (d) 
      at the same mass imbalance $D_f/D_c = 0.4$ in a smaller frequency range.
      Panels (a), (c) and (d) show two metallic (insulating) solutions, one in 
      the purely metallic (insulating) phase
      (blue (red)) and one within the coexistence 
      region (cyan (orange)). The precise $U$ values of these points 
      are given in the text.} 
\end{figure}

Fig.\ \ref{fig2:siw} shows this imaginary part of the self-energy for
exemplary $U$ values at $\beta=100$. 
In panel (a) the results for the HM are shown: Specifically,
one metallic  (M, $U=4.6$, blue)
and one insulating solution (I, $U=4.9$, orange)
upon decreasing $U$ ($\leftarrow$), as well as
one M ($U=5.2$, cyan)          
and I solution ($U=5.5$, red) upon increasing $U$ ($\rightarrow$).
The insulating solution is characterized by a divergence 
$\Sigma(\text{i}\omega_n)\rightarrow - \infty$ for $\omega_n\rightarrow 0$,
and  we take this as the criterion for discriminating M and I solutions in the phase diagram 
(Fig.\ \ref{fig1:PD}).

Fig.\ \ref{fig2:siw} (b) displays the results for the FKM:
Analogously, one metallic solution ($U=1.5 <U_\text{c}$, blue) and
one insulating solution ($U=2.5>U_\text{c}$, red) 
are shown. Since there is no coexistence region in the FKM, we instead additionally show the 
solution for the unique and temperature-independent $U_\text{c}=2$ (yellow).

Fig.\ \ref{fig2:siw} (c,d) display analogous phase points
though now for the mass-imbalanced HM at a 
mass imbalance $D_f/D_c=0.4$, which is inbetween the 
the limiting cases of the HM (a) and FKM (b).  
The more mobile $c$ fermions are shown in Fig.\ \ref{fig2:siw} (c), and
the less mobile $f$ fermions in (d).
The shift of the Mott-insulator transition compared to the HM 
is taken into account by adjusting the interaction strength to 
$U=3.2$ ($\leftarrow$ M, blue), $U=3.3$ ($\leftarrow$ I, orange), $U=3.4$ 
($\rightarrow$ M, cyan) and $U=3.5$ ($\rightarrow$ I, red) in the two lower
panels.
We observe that the $c$ fermions undergo a stronger renormalization than
the already less mobile, and thus heavier, $f$ fermions.
That is, in the metallic phase the effective mass enhancement  
\begin{equation}
m^\ast/m=Z^{-1}=1- \left.\partial \text{Im} \Sigma(\iw_n)/{\partial \iw_n}\right|_{\iw_n \to 0}
\label{Eq:Z}
\end{equation}
is stronger for the $c$ fermions. Nonetheless, the divergent 
$\text{Im} \Sigma(\iw_n)\sim -1/(\iw_n)$ behavior  of the self-energy which
indicates the insulating phase sets in at the same $U$ value. Overall both
self-energies in (c,d) look similar as for the HM in (a) except for the smaller
$\iw_n$ ($x$-axis) scale, which can be understood from the lower $U$ value of the Mott transition.

\begin{figure}
  \includegraphics[width=\columnwidth]{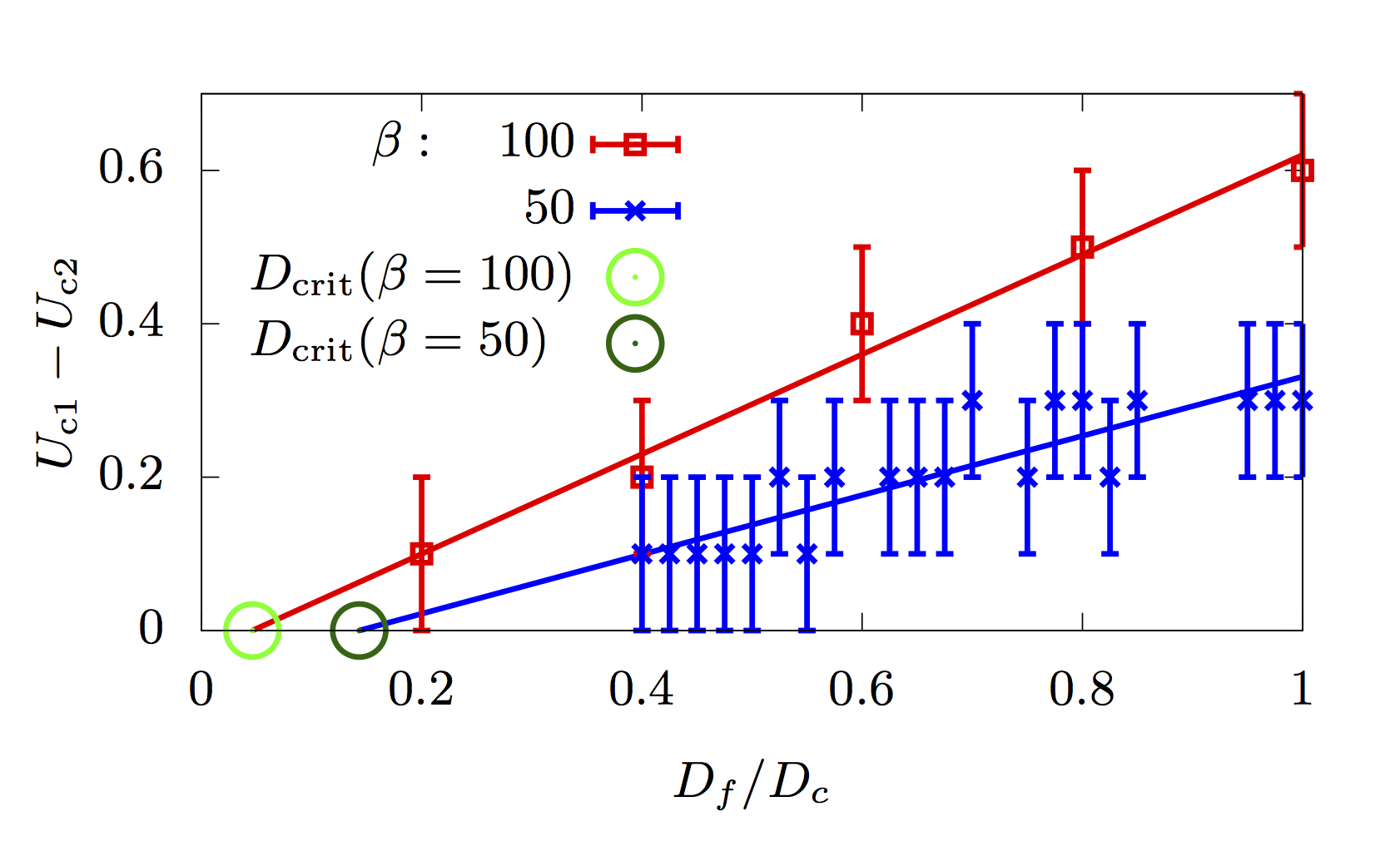}
	\caption{\label{fig5:coex} Difference of the
	critical interaction strengths, $U_\text{c1}-U_\text{c2}$, 
    of the coexistence region
    vs.\ the mass imbalance $D_f/D_c$.
	The linear extrapolation yields the temperature-dependent 
    critical point $D_\text{crit}$ at which the first order 
    transition with coexistence region ends.
      	}   
\end{figure}

In Fig.\ \ref{fig5:coex}, we  extrapolate the coexistence
region which is shrinking with decreasing $D_f/D_c$ to
$U_\text{c1}-U_\text{c2}=0$. This marks the temperature-dependent 
critical point $D_\text{crit}(\beta)$ at which the 
first order transition ends,
denoted by a green circle in Fig.\ \ref{fig5:coex}. 
At the lower temperature ($\beta=100$) this critical point
is already very close to the FKM limit $D_f/D_c=0$, which 
together with the apparent temperature dependence 
suggests that indeed for zero temperature 
$D_\text{crit}(\beta=\infty)=0$. In other words, the 
FKM appears to be the critical point of the mass-imbalanced 
HM at zero temperature.

\begin{figure}
  \includegraphics[width=\columnwidth]{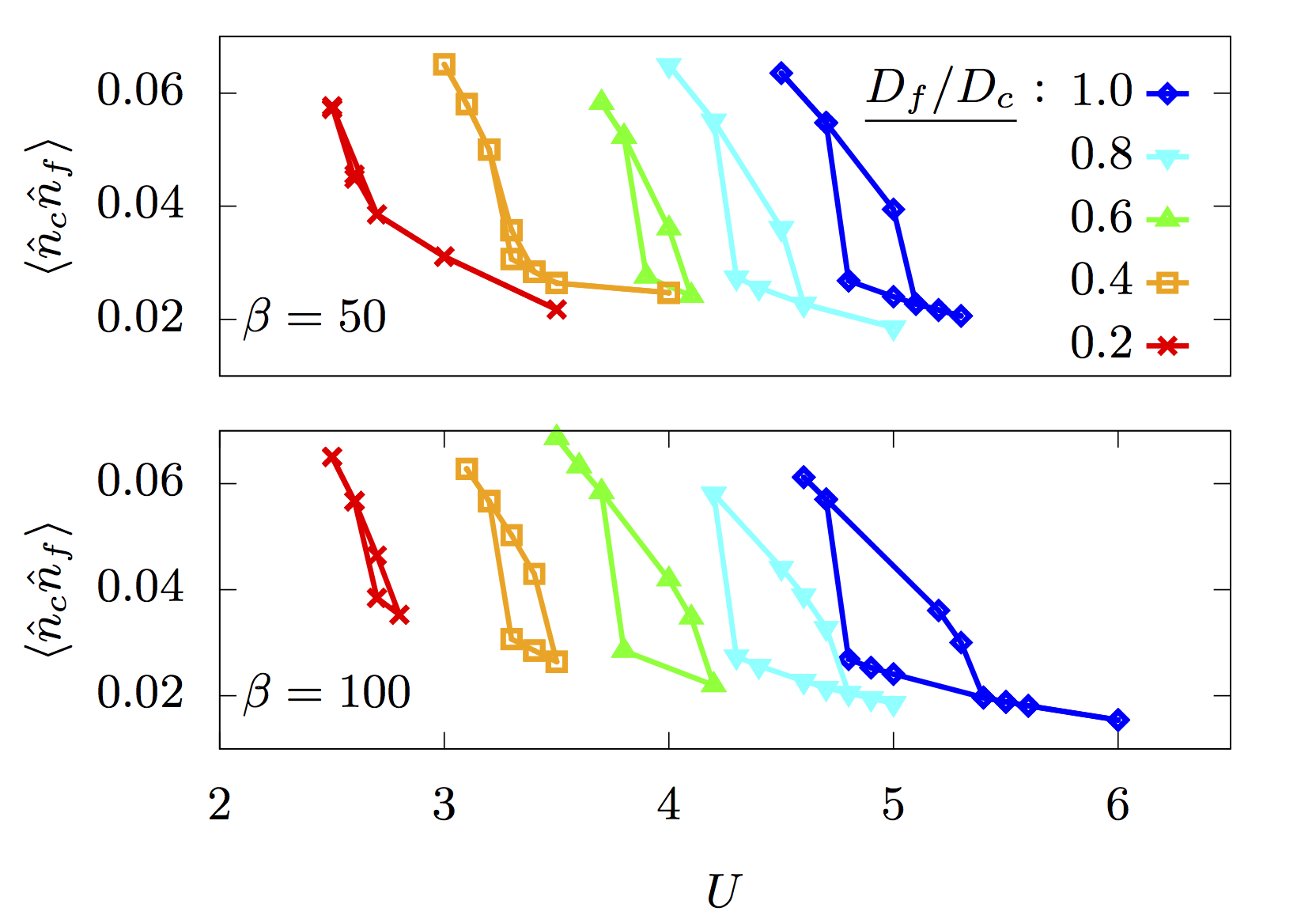}
	\caption{\label{fig4:double occ} Double occupancy 
	$\langle \hat n_{c} \hat n_{f} \rangle$ for
	different values of the mass imbalance $D_f/D_c$
	at $\beta=50$ (upper panel) and $\beta=100$ (lower
    panel), showing the hysteresis loop of the first order 
    transition upon increasing and decreasing $U$.
   	}
\end{figure}

The double occupancy in
Fig.\ \ref{fig4:double occ} likewise shows the first order transition 
and hysteresis loop in $U$. In the Mott-insulating phase each site 
is essentially singly occupied. 
In contrast to when approaching the 
metallic phase upon decreasing $U$,
the higher kinetic energy leads to a strong increase in the number of 
doubly occupied sites. 
Only in the limit $U \rightarrow 0$ the 
uncorrelated double occupation $\langle \hat n_{c} \hat n_{f} \rangle=1/4$ 
is eventually recovered.
The figure shows again clearly that the coexistence region increases 
upon decreasing temperature (increasing $\beta$). On that ground the behavior 
of the mass-imbalanced HM is similar to the HM also regarding the double occupation.

\begin{figure}
  \includegraphics[width=\columnwidth]{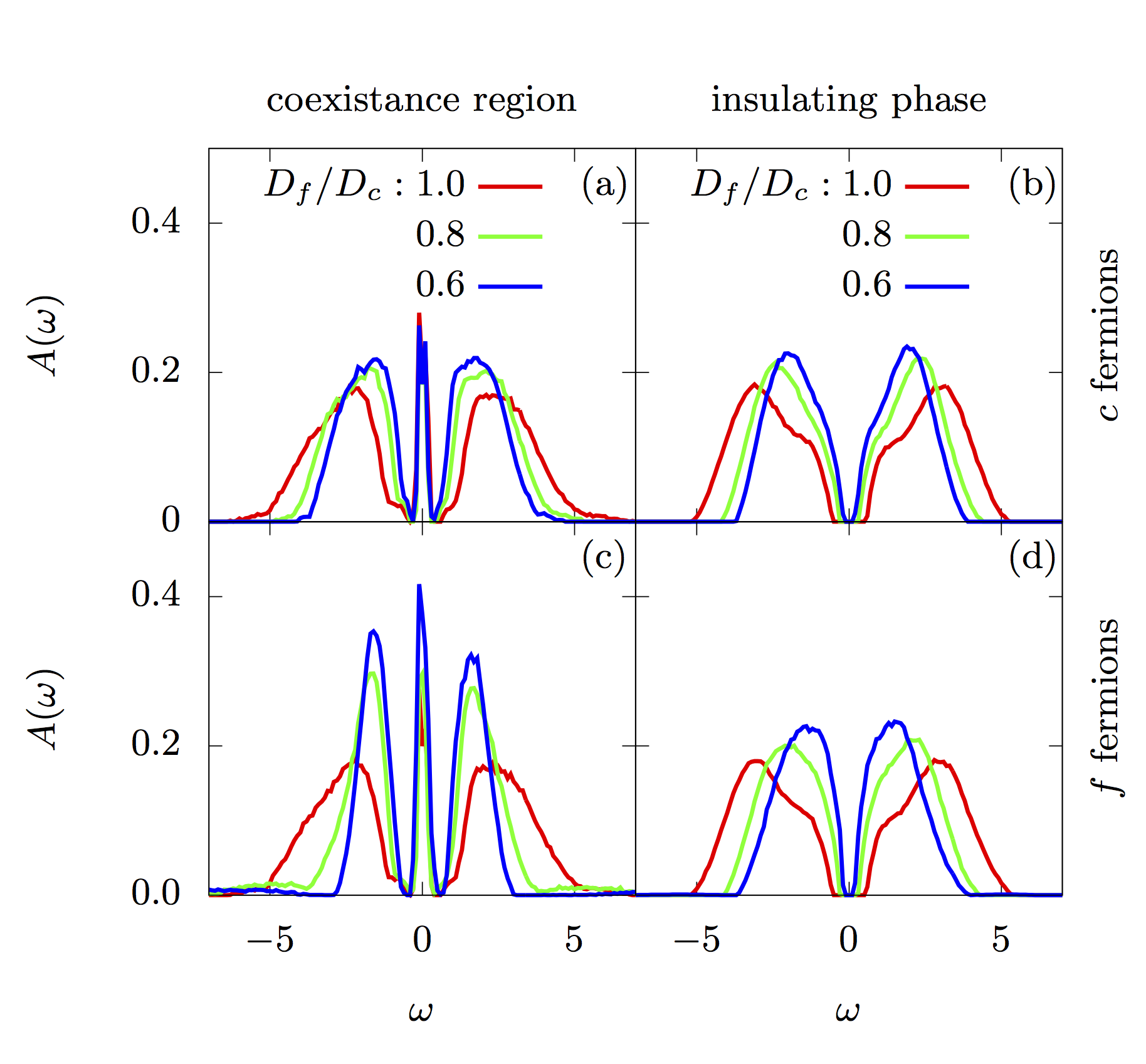}
          \caption{\label{fig7:spectra} Spectral function $A(\omega)$ near
          the critical interaction strength $U_\text{c1}$ for different
          mass imbalances $D_f/D_c$ at $\beta=100$. 
          The upper and lower two panels show  the more and less
          mobile  $c$ and $f$ fermions, respectively.
          The left (right) panels
          show a metallic (insulating) solution just below (above) $U_\text{c1}$, indicated by  the three vertical arrows in 
          Fig.\ \ref{fig1:PD}. Specifically the $D_f$, $U$ values for the left panels
          (a) and (c) are: $D_f=2.0$, $U=5.2$ (red); $D_f=1.6$, $U=4.1$ (green);
          $D_f=1.2$, $U=3.4$ (blue); and for the right panels
          (b) and (d): $D_f=2.0$, $U=5.5$ (red); $D_f=1.6$ $U=4.2$ (green),
        $D_f=1.2$ $U=3.5$ (blue).}        
\end{figure}

Fig.\ \ref{fig7:spectra} shows the corresponding spectral function
comparing the HM ($D_f/D_c=1$) and the mass imbalanced HM 
($D_f/D_c=0.8$ and 0.6), immediately before (a,c) and after 
(b,d) the transition $U_\text{c1}$ where the metallic 
solution ceases to exist.
We see that the spectral functions are actually quite similar
with a three-peak structure on the metallic side, consisting of a 
lower and upper Hubbard band and a central quasiparticle peak inbetween. 
Immediately after the transition $U_\text{c1}$ there is a 
gap in the insulating phase. This is very different from the FKM in 
Fig.\ \ref{fig0:scheme} where we have two peaks which do (do not) not
overlap any more for $U$ below (above) the transition. 
There is also no indication of a smooth crossover in 
Fig.\ \ref{fig7:spectra} from the HM behavior
to that of the FKM. The major difference is that with reducing $D_f/D_c$ 
in  Fig.\ \ref{fig7:spectra}, the position of the upper (lower) Hubbard band
shift to higher (lower) energies, which is in agreement with the reduced
$U_\text{c1}$ value at which the transition occurs.

\section{Discussion}
\label{Sec:discussion}
The physics  in Fig.\ \ref{fig7:spectra} of the mass-imbalanced HM seems to resemble that of the HM rather than that of the FKM. Also in   Fig.\ \ref{fig5:coex} we have seen that the first order transition
of the HM survives a quite considerable mass imbalance, down to even smaller  $D_f/D_c$ values at lower temperatures.
We know on the other hand that there is no first order transition
for the FKM. This poses the questions whether at
zero temperature $D_\text{crit}=0$, i.e., there is always a first order transition for $D_f>0$? Is  the  FKM and its physics  a singular point of the mass-imbalanced Hubbard model at zero temperature?
Indeed the temperature dependence in 
 Fig.\ \ref{fig5:coex} already suggests  $D_\text{crit}=0$.

Let us further address this question by considering the underlying DMFT impurity problem which will help us to identify the fundamental physics. In the case of the HM, DMFT corresponds to the self-consistent solution of the Anderson impurity model,\cite{Georges96a} whereas for the FKM one needs to iterate the resonant level model.\cite{Freericks03} In case of the mass-imbalanced HM, we have 
an Anderson impurity type of model but with a different bath for the more mobile $c$ and the less mobile $f$ fermions.

Specifically the DMFT impurity problem of the mass-imbalanced HM has, in case of the Bethe lattice,
the following  non-interacting Green function 
\begin{equation}
{\cal G}^{-1}_x(\omega) = \omega + \mu -(D_x/2)^2 \; G_x(\omega) \equiv \omega + \mu -\Delta_x(\omega),
\label{Eq:Bethe}
\end{equation}
 given by the interacting Green function 
$G_x(\omega)$ at frequency $\omega$.\cite{Georges96a} This can be understood as follows:
if a fermion leaves the impurity  with amplitude $D_x/2$ it propagates through the strongly correlated bath of the other sites given by   $G_x(\omega)$ before
returning to the impurity  with amplitude $D_x/2$.
The difference for the mass-imbalanced HM is now that each fermion species $x=c,f$ has an individual bath or hybridization function $\Delta_x(\omega)$. 

On the metallic side, close to the metal-insulator transition we have
seen a narrow central resonance in Fig.\ \ref{fig7:spectra}, and we can map the Anderson impurity model onto a Kondo model with different baths. For a narrow resonance, we are in the Kondo regime and can apply the  perturbative renormalization group also known as  Anderson's poor man scaling\cite{Anderson,Hewson} for understanding the physics.
For the renormalization of the Kondo coupling, 
the spectral weight at high energies, i.e., in the upper and lower Hubbard band does not matter\cite{footnotepoorman} and we can restrict ourselves to the central resonance.\cite{Held13} 

Since the Kondo effect relies on spin-flip (i.e., $c$-$f$) scattering, \cite{Hewson} it is not possible that the coupling to the bath for one species alone becomes large in the renormalization process
whereas the other does not. The two fermionic species make the Kondo effect at essentially the same Kondo temperature and develop a Kondo resonance together. This also implies that the physical origin of the central resonance and its width is the same for  $c$ and $f$ fermions in Fig.\ \ref{fig7:spectra}. The height of the central resonance, on the other hand, is independent of 
$D_f$  for the $c$ fermions (note $D_c=2$ is fixed), whereas it increases  $\sim 1/D_f$ for the less mobile $f$ fermions as does the non-interacting DOS of the  $f$ fermions.

We can validate this expectation numerically  by estimating the 
width of the central resonance as  $Z_xD_x$,
where $Z_x$ is the quasiparticle renormalization as calculated  from the self-energy through Eq.\ (\ref{Eq:Z}).
If the width $Z_xD_x$  is to be the same for the  $x=c,f$ fermions, 
the quasiparticle renormalization  $Z_f$ needs to be larger since $D_f<D_c$.
Indeed a first indications for this we have already seen in the smaller $f$ self-energy and hence larger  $Z_f$ in Fig.\ \ref{fig2:siw}.

\begin{figure}
    \includegraphics[width=\columnwidth]{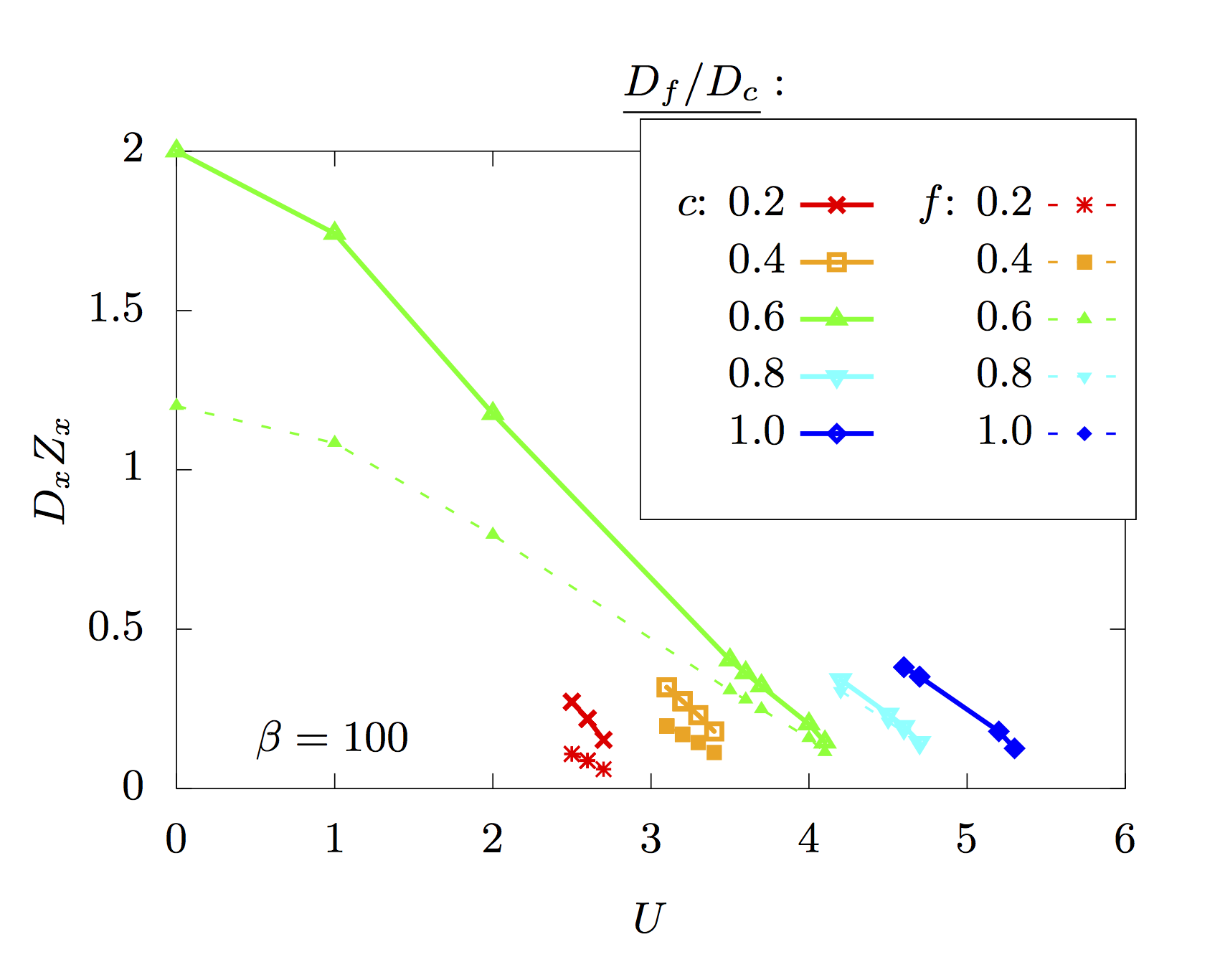}
	\caption{\label{fig6:ZD} 
        Width of the central quasiparticle peak $Z_x D_x$ for the 
        fermionic species $x=c,f$ (solid line, dashed line) 
        vs.\ $U$ at $\beta=100$ and different values of the 
        mass imbalance $D_f/D_c$.
	       }  
\end{figure}

 Fig.\  \ref{fig6:ZD} shows the analysis of the  quasiparticle renormalized width
 $Z_xD_x$ for the two fermionic species $x$. Indeed we see that $Z_cD_c \rightarrow  Z_fD_f$ as the metal-insulator transition is approached, in agreement with our considerations above. 
In contrast, at $U=0$  where $Z_x=1$ the width is obviously very different. For the exemplary green line in Fig.\  \ref{fig6:ZD}, we see the continuous evolution
from a factor of $0.6$ difference in  width
 at $U=0$ to the same width in the Kondo regime in the vicinity of the phase transition line  $U_\text{c1}$.

The observation that $Z_cD_c \rightarrow  Z_fD_f$ for $U\rightarrow U_\text{c1}$ also implies a simultaneous phase transition for both fermionic species, in agreement with our observation
for the  self-energy in Fig.\ \ref{fig2:siw} and  the phase diagram Fig.\ \ref{fig1:PD} above. This is in contrast to the multi-orbital HM for which an orbital-selective phase transition has been observed.\cite{Liebsch03b,Biermann05b,Koga04a,Arita05a,Knecht05a,Ferrero05a} The striking difference is that for the multi-orbital HM we have two spin species for each orbital
so that the Kondo effect can occur separately for each orbital which is not possible in our case of the mass-imbalanced HM.

Altogether, the physics of the mass-imbalanced HM  for $D_f>0$ rather resembles that of the HM than that of the FKM. Noteworthy,  however, one needs ever lower temperatures with decreasing $D_f$ in order for the central resonance to develop since the joint Kondo temperature decreases exponentially. The similar physics to the HM further suggest the same first order metal-insulator transition.
Indeed the temperature dependence of the critical point in  Fig.\ \ref{fig5:coex}
already suggested  that the transition is of first order for any $D_f>0$ at zero temperature. Let us note however that while the similar physics suggests the
same first-order nature of the phase transition, in principal there could also be a continuous transition in the HM given the underlying Kondo physics. 
Indeed whether the phase transition is of first order in the HM was quite debated in the early days of DMFT\cite{Georges92,Schlipf99,Rozenberg99} and eventually had to be decided numerically.
In our case of the mass-imbalanced HM our numerical data strongly indicates that the phase transition already  becomes
first order as soon as there is a finite hopping of the $f$ fermions ($D_f>0$).

\section{Conclusion}
\label{Sec:conclusion}
We have analyzed the Mott-insulator transition of the mass-imbalanced Hubbard model within the paramagnetic phase. Our phase diagram, Fig.\  \ref{fig1:PD}, shows a first order phase transition in a wide range of mass imbalances $D_f/D_c$. With decreasing temperature the region of first order coexistence expands; and our results suggest that the mass-imbalanced Hubbard model always displays a first order metal-insulator transition at zero temperature as soon as a small, but finite, hopping of the less mobile $f$ fermions is switched on ($D_f>0$).

For the FKM ($D_f=0$), we have two bands along with a gap opening with increasing $U$ as soon as these two bands do not overlap any longer. If we switch on $f$ fermion hopping however ($D_f>0$), a central resonance in this gap develops due to the Kondo effect. This stabilizes the metallic phase and shifts the metal-insulator transition in the phase diagram  Fig.\ \ref{fig1:PD} towards larger $U$ values. 
This resonance and a three-peak structure can be seen in the spectral function,
Fig.\ \ref{fig7:spectra}. We find that the width of the
central resonance is the same  for the $c$ and $f$ fermions.
This can be understood from the fact that the spin-flip (here $c$-$f$) scattering 
is crucial for the Kondo effect. Hence we have a joint Kondo temperature and width 
of the Kondo resonance for $c$ and $f$ fermions. 

This is affirmed by an analysis of the quasiparticle renormalization factor $Z_x$ in
 Fig.\  \ref{fig6:ZD}, which shows  $Z_cD_c \rightarrow  Z_fD_f$  when approaching the metal-insulator transition, i.e. when we are in the Kondo regime accompanied by a narrow central resonance. It further shows that the metal-insulator transition occurs simultaneously for both, $c$ and $f$, fermions in 
the mass-imbalanced Hubbard model. The Falicov-Kimball physics, 
where the $f$ fermions are insulating for any $U$ and the $c$ fermions for $U>D_c$, 
is a singular point of the phase diagram at zero temperature.

Altogether our results show that the physics of the mass-imbalanced Hubbard model
in the paramagnetic phase resembles that of the Hubbard model. This is because of 
the equalizing power of the joint Kondo effect of the two fermionic species. 
Regarding the antiferromagnetic phase we nonetheless expect a qualitatively different behavior: 
the mass imbalance breaks the $c$-$f$ O(3) rotational symmetry of the order parameter;
and Monte-Carlo simulations\cite{Liu15} indeed indicate an Ising-type ordering. Hence we 
expect Ising-type critical exponents similar to what has recently been reported 
for the FKM,\cite{Antipov14} whereas we have a Heisenberg-type of ordering and associated 
critical exponents\cite{Rohringer11} for the Hubbard model.

\section{Acknowledgment}
We thank B. Hartl, C. Taranto, P. Thunstr\"om,  A. Toschi, T. Ribic, and V. Zlati\'c 
for useful discussions, as well as A. Sandvik for making available his maximum entropy program. 
This work has been supported  by European Research Council under the European Union's Seventh
Framework Program (FP/2007-2013)/ERC through grant agreement n.\ 306447; the  Vienna
Scientific  Cluster  (VSC)  Research  Center  funded  by  the
Austrian Federal Ministry of Science, Research and Economy
(bmwfw); and the Austrian Science Fund (FWF) through  SFB ViCoM F41  and  project  I-610-N16 as part of the research unit FOR 1346 of the Deutsche Forschungsgemeinschaft  (DFG). 
Calculations have been done on the Vienna Scientific Cluster~(VSC).


\begin{thebibliography}{}
\bibitem{footnoteschematicMIT} DMFT spectra for the FalicovKimball model have been calculated analytically at $U=0, 1.5, 2.0, 5.2 ; D=2$; those of the Hubbard model have been calculated at $U=0, 4.0, 5.2, 5.5 ; D=2$ using continuous quantum Monte Carlo simulations as a DMFT impurity solver and the maximum entropy method.
\bibitem{footnotepoorman} At least as long as these Hubbard bands are well enough separated from the central resonance.
\end{thebibliography}
\end{document}